# Water jet space charge spectroscopy: Route to direct measurement of electron dynamics for organic systems in their natural environment.

Michael Mittermair,[1] Felix Martin,[1] Martin Wörle,[1] Dana Bloß,[2] Andreas Duensing,[1] Reinhard Kienberger,[1] Andreas Hans,[2] Hristo Iglev,[1]* André Knie,[2]* Wolfram Helml,[3]*

[1]Physik Department E11, Technische Universität München, James-Franck-Str. 1, D-85748 Garching, Germany.
[2]Institut für Physik und CINSaT, Universität Kassel, Heinrich-Plett-Straße 40, D-34132 Kassel, Germany
[3]Technische Universität Dortmund, Maria-Goeppert-Mayer-Str. 2, D-44227 Dortmund, Germany.
*Corresponding author, Email: higlev@ph.tum.de, knie@physik.uni-kassel.de, wolfram.helml@tu-dortmund.de

**Abstract**
The toolbox for time-resolved direct measurements of electron dynamics covers a variety of methods. Since the experimental effort is increasing rapidly with achievable time resolution, there is an urge for simple and robust measurement techniques. Within this paper prove of concept experiments and numerical simulations are utilized to investigate the applicability of a new setup for the generation of ultrashort electron pulses in the energy range of 300 eV up to 1.6 keV. The experimental approach combines an in-vacuum liquid microjet and a few-cycle femtosecond laser system, while the threshold for electron impact ionization serves as a gate for the effective electron pulse duration. The experiments prove that electrons in the keV regime are accessible and that the electron spectrum can be easily tuned by laser intensity and focal position alignment with respect to the water jet. Numerical simulations show that a sub-picosecond temporal resolution is achievable.

**Introduction**
Over the past years the concepts for the observation of fundamental processes connected with electron reconfiguration dynamics in real time have been developed continuously (*1–3*). This would lead to a more direct understanding of changes in physical systems based on the underlying electron configuration and open the opportunity to generate and control desired properties and reaction pathways in a deliberate and efficient manner (*4–7*). Naturally, for these pioneering experiments the objects of investigation have been chosen to be as simple as possible to demonstrate the potential of new ultrafast methods and fully develop their capabilities in a well-understood setting. Thus, while the temporal resolution has been improved from picoseconds to femtoseconds and attoseconds, the systems under investigation were continuously becoming less complex - from chemical reactions in molecules and transient dynamics at metal surfaces to excitation and relaxation processes in the most basic systems like the helium (*8*) and hydrogen atom (*9–11*).
Now that our understanding of these techniques has ripened and grown, we are on the brink of broadening the range of ultrafast research again to practically more relevant subjects like the internal charge separation in solar cells (*12, 13*), efficient current control in electronic materials (*14, 15*) or the bondbreaking and re-structuring in biomolecules (*16, 17*).
An area of very high interest is the complex of open questions surrounding damage of DNA molecules following UV radiation or electron impact (*18–20*). This has consequences for cancer radiation therapy (*21, 22*), the problem of skin burn by sun light and the effects of secondary radiation on organic cells in a technologically quickly evolving world (*23, 24*).
Since all those biologically relevant reactions occur within human cells and thereby basically in liquid solution, there is an undeniable need for a robust tabletop technique, which is capable of probing organic molecules in a liquid environment while maintaining ultrashort time resolution. We want to pave the road for a new technique that satisfies these needs.

**Results**
**A new method for generating ultrashort electron pulses**
In solid-state photoemission experiments the so-called space-charge effect, where the repellent Coulomb force between the electrons escaping from the surface is an unwanted but permanent companion (*25*). While such a cloud of electrons is obstructive in photoemission experiments, we see a chance to exploit it as an ultrashort source of electrons with kinetic energies in the keV region. We propose a novel experimental apparatus consisting of a liquid microjet combined with a titanium-sapphire laser system which delivers bandwidth limited few-cycle laser pulses. These ultrashort laser pulses generate a cloud of free electrons in the gaseous vicinity of the liquid jet as schematically shown in Figure 1. The electrons within the cloud experience repulsion due to the Coulombic interaction, which eventually leads to expansion of the cloud. While the inner electrons feel the repelling force from all sides, which does not allow them to gain much kinetic energy, the outer ones feel the pushing force from the cloud and gain the highest kinetic energy. These fastest electrons, which can be referred to as cutoff electrons build an expanding shell which is followed by a trail of slower electrons. In 1991, Ammosov derived a straightforward formula to describe the behavior of the fastest electrons by applying the energy conservation law for an expanding sphere of electrons (*26*):

$$\frac{1}{4\pi\varepsilon_0}\frac{Ne^2}{r_0} = \frac{1}{4\pi\varepsilon_0}\frac{Ne^2}{r} + \frac{m_e v^2}{2} \quad (1)$$

with the electron mass $m_e$, charge $e$ and electron velocity $v$. $N$ is the number of uniformly distributed electrons inside a sphere of radius $r_0$ generated at $t = 0$. The electron kinetic energy for very large radii of the expanding shell ($r \to \infty$) is given by $E_{kin} = Ne^2/4\pi\varepsilon_0 r_0$. These electrons will be emitted in all directions from their place of origin. In particular, a certain quota of the fast electrons will hit the liquid jet and interact with water molecules or any species of organic compound dissolved inside the liquid. The basic idea behind the experiment is to use the fastest electrons, which inherently emerge within an extremely narrow time window, as pump or probe in a time-resolved experiment. As only the fastest electrons have sufficient energy to ionize a core level electron, the element specific threshold for electron impact ionization can be used to distinguish between fast and slow ones. By keeping the high energy shoulder of the electron spectrum just above the threshold an electron pulse with an ultra-short effective pulse duration can be achieved. Biologically relevant molecules mostly consist of carbon, nitrogen and oxygen with their K-shell ionization thresholds at 296 eV, 406 eV and 543 eV, respectively (*27–29*).

**Proof-of-principle experiments**
To establish the feasibility of this new approach it is required to clarify two aspects. Foremost, one needs to show that the electron acceleration by the charged cloud is strong enough to create electrons with a kinetic energy higher than the respective element specific threshold for electron impact ionization. Secondly, scrutinizing the time structure of the electron bunch is necessary to demonstrate the applicability of this approach for ultrafast spectroscopy experiments. From equation (1) we know that the final kinetic energy of the electrons is linked to the expansion of the charge cloud. Therefore, it is obligatory to examine the relation between the initial distance between the electron cloud and the liquid microjet, which sets the effective radius of the expanded charge cloud. The distance between laser focus and liquid jet is varied by the manipulator system keeping all other parameters constant. Figure 2 shows how the kinetic energy cutoff, which is defined as the energy where the signal intensity drops to 30% of its plateau value, changes with the distance between the liquid microjet and the laser focus. With the cutoff energy being proportional to the number of generated free electrons as shown in equation (1), it is obviously proportional to the number of molecules at the laser focus and thereby to the gas pressure. According to Faubel et al. (*30*), the radial pressure dependency in a liquid microjet experiment follows the rule for a cylindrical particle source

$$p(R) = \frac{R_0}{R} p_0 \text{ for } (R > R_0) \tag{2}$$

with radial distance $R$, the equilibrium pressure at the liquid surface $p_0 = 23.4$ mbar at a liquid temperature of 20 °C and the jet radius $R_0$ of 12.5 µm. It is possible to fit

$$E_{cutoff}(R) = \frac{A}{R - R_C} + B \tag{3}$$

to the data. Here, the center of the liquid jet is represented by $R_C$. $A$ is a constant scaling factor, which takes into account the size of the initial electron cloud and the ionization probability per water molecule. In general the effective density of free electrons depends on various parameters like the ionization cross section and laser intensity as well as the evaporation rate of the microjet and the initial size of the laser focus. From comparison of equation (3) and equation (2) it follows that $E_{cutoff} \propto p(R) + B$ within the observed pressure range. The parameter $B$ is a linear offset, which is needed for the fit to match the data. Physically it might originate from electrons which have been accelerated in the laser field instead of acceleration by the charged electron cloud. From the known distance between laser focus and jet it is possible to determine the gas pressure for every measured spectrum, following the red line in Figure 2. One can see that the electrons reach kinetic energies in the keV regime which is sufficiently high to introduce inner shell vacancies (K-shell) in light elements up to magnesium. Notably, these elements play an extraordinary role when it comes to biological relevant molecules. The initial electron cloud is generated via laser induced ionization of the water vapor. For high laser intensities the distinction between multiphoton and tunneling/barrier suppression ionization is given by the Keldysh adiabaticity parameter (*31*)

$$\gamma = \omega \frac{\sqrt{2 m_e I_p}}{e \mathrm{E}}, \tag{4}$$

with the frequency of the driving laser $\omega$, the ionization potential $I_p = 12.6$ eV of the 1b$_1$ molecular orbital of water (*32, 33*) and the laser field strength $\mathcal{E}$. The maximum intensity calculated from the applied laser parameters and focus conditions reaches up to $1.3 \times 10^{17}$ W/cm$^2$, which corresponds to a maximum field strength $\mathcal{E}_{max} = 1 \times 10^{12}$ V/m, leading to $\gamma = 0.033$, a clear indicator that tunneling or barrier suppression ionization is dominant. These regimes are separated by the barrier-suppression field strength $\mathcal{E}_{bsi}$ which marks the condition where the atomic ionization potential and the effective one-dimensional potential barrier are equal assuming a constant electric field (*34*)

$$\mathcal{E}_{bsi} = \frac{I_p^2}{4Z} \tag{5}$$

where $Z$ is the charge of the remaining ion (*35*). Thus, with the electric field $\mathcal{E}_{max}$ clearly exceeding the barrier suppression limit $\mathcal{E}_{bsi} = 5.3 \times 10^{10}$ V/m for the weakest bound electron multiple electron ionization is likely to occur.

**Modelling space charge acceleration**
To reach applicability in ultrafast science it is necessary to examine the time structure of the electron bunches and to validate that the electrons reach those energies before they hit the liquid jet surface. For this purpose we employ a two stage Monte Carlo simulation. A closer description of the simulation can be found in the methods section. Figure 3 compares the simulation results with experimental data. The solid lines show electron spectra for laser intensities of $2.6 \times 10^{16}$ W/cm$^2$ (blue), $4.5 \times 10^{16}$ W/cm$^2$ (orange), and $5.3 \times 10^{16}$ W/cm$^2$ (green). One can observe how the spectra shift towards higher energies with increasing laser intensity. The dashed lines represent the results from a set of simulations that only differ in the total number of ionized electrons $N$. The simulations are selected to fit the high energy cutoff of the corresponding experimental spectra with the kinetic energy being calculated from the velocity component pointing towards the detector. We find excellent agreement in the high-energy cutoff. The slow electrons, which do not have sufficient energy to liberate an electron from an inner shell, deviate from the quantitative prediction and are not part of the further considerations. To classify the cutoff energies achieved, the K-shell binding energies of selected elements are plotted on the upper abscissa in Figure 3. For demonstration, we take a closer look onto the 300 million particle simulation (blue dashed line in Figure 3). Figure 4 A shows the energy and spatial evolution of the simulated electron cloud for a series of instances in time. Again, we only take into account the contribution to the kinetic energy by the velocity in the direction towards the detector. The purple line shows a fit of the fastest electrons by

$$E_{kin}(r) = \frac{Ne^2}{4\pi\varepsilon_0}\left(\frac{1}{r_0} - \frac{1}{r}\right) \tag{6}$$

which is a rearranged version of equation (1). One can see that the electrons gain most of their kinetic energy within the first 300 µm. The dashed lines mark the K-shell ionization thresholds for carbon (red), nitrogen (blue), and oxygen (green). Figure 4 B shows the derived effective electron pulse duration versus the distance to the center of the electron cloud. The effective duration is calculated as the full-width at half maximum of the temporal distribution for electrons with kinetic energy above the respective K-shell ionization threshold. The pulse shape linearly disperses with distance, whereby a threshold closer to the electron cutoff allows for smaller dispersion as well as for shorter pulse duration.

**Discussion**
As noted in the introduction, there is a growing demand for investigating organic molecules in their natural liquid environment with an ultrashort time resolution. Obviously, it is a significant step to reduce the requirements for these measurements usually conducted in large-scale facilities to a tabletop technique. We tackle this by combining an in-vacuum liquid microjet with an ultrashort-pulse laser system, capitalizing on the ever present space charge effect. The extremely small distance between the origin of the electrons and their experimental utilization allows the electron pulse to stay confined during acceleration. As driving mechanism for the electron acceleration the Coulombic repulsion due to space charge within a cloud of electrons is exploited. The necessary ultrashort electron pulse duration is achieved by utilizing the threshold for electron impact ionization of inner shells in the target elements in combination with time structure of the electron cloud as temporal gate.

Varying the distance between laser focus and liquid microjet provides a parameter that can be used to efficiently tune the resulting electron spectrum. With kinetic energies in the low keV regime we demonstrated that electron impact ionization of core holes is accessible, which is the most important necessity that must be met to allow this technique to work. The applied numerical model reproduces the experimental data remarkably well, which implies that the simulation depicts the acceleration mechanism sufficiently in detail. The extracted temporal and spatial evolution of the fast electrons shows that the electrons gain most of their kinetic energy within a very short distance from their origin. The effective electron pulse duration is calculated for a set of exemplary atomic species, demonstrating that a duration shorter than one picosecond can be achieved.

Our work lays the foundation for a new experimental technique relying on space-charge-driven electrons as a direct in-situ excitation source for ultrafast measurements in liquids. To complete the scheme also a mechanism suited for probing the ionized molecules has to be implemented. Either the fluorescence rate or the time-dependent Auger electron yield would serve as an element-specific probe signal. Thus, a versatile, affordable and compact experimental setup is made available for studies on biologically relevant systems in their natural environment.

**Materials and Methods**
**Laser system**
All measurements were performed with a commercial master oscillator (*Femtosource™ Rainbow™*) power amplifier (*Femtopower™HEHR CEP4*) system. The pulses were spectrally broadened in a home-built differentially pumped hollow core fiber setup and temporally compressed within a chirped-mirror array. The pulse length was optimized for maximum cutoff energy. This laser system delivers pulses with an energy of up to 1.5 mJ and a pulse length shorter than five femtoseconds at a repetition rate of 4 kHz. The laser wavelength is centered at 690 nm with a linear polarization. A more detailed description of the system can be found in (*36*).
**Liquid microjet**
The liquid microjet was produced by injecting water with a high-performance liquid chromatography (HPLC) pump through a 25 μm sized glass nozzle into the vacuum. A flow of 0.8 ml/min required 8 bar backing pressure. Adjustment by a 3D~manipulator system allowed for fine tuning of the distance between water jet and laser pathway. After interaction with the laser beam, the jet was collected by a cryogenic cold trap, cooled with liquid nitrogen from outside the vacuum. With a second cold trap in the interaction chamber a pressure of $10^{-2}$ mbar in the interaction region and $10^{-5}$ mbar in the electron spectrometer was maintained, ensuring proper operation of the electron detector. A depiction of the vacuum setup can be found in the supplement. The jet was operated at room temperature and grounded outside the vacuum via a gold wire connected to the liquid near the vacuum feedthrough to prevent charging of the liquid and the nozzle. A low concentration (50 mM max.) of NaCl was added to the water solution for the same reason.

**Time-of-Flight Spectrometer**

All spectra were recorded with an *ETF11* electron time-of-flight spectrometer from *Stefan Käsdorf Geräte für Forschung und Industrie*. The spectrometer is designed to work under high vacuum conditions below $10^{-6}$ mbar and has an entrance aperture of 3 mm. With the backing pressure in a liquid microjet experiment being way higher than in solid state experiments, it was necessary to reduce this aperture to ensure proper vacuum conditions inside the drift tube. This is achieved by implementing a custom made 170 μm pinhole which reduces the aperture and thereby the gas flow by factor of about 400. Electrons are measured in a direction perpendicular to the laser polarization. For readout a *Teledyne LeCroy Waverunner 104 MXI* oscilloscope is used.

**Simulation**

In order to link the experimental results with the underlying mechanisms, a Monte Carlo simulation based on classical transport theory was implemented. It includes the electron acceleration by the electric field of the laser as well as the influence of the space charge effect due to the high density of generated free electrons. The simulation was divided into two consecutive parts: the first part handles the ionization process and the electron acceleration in the electric field of the laser pulse. It also takes into account elastic and inelastic scattering events with the surrounding water molecules. The second part addresses the space charge effect harnessing the charge distribution from the first part as a starting point.

**Ionization and acceleration in the electric field**

The time dependent electron release profile at high laser intensities as they are employed in the experiment can be estimated using the ADK-theory described by Corkum et al. (*37*):

$$\Gamma(t) \propto \frac{1}{E_0 f(t)} \exp\left[-\frac{2}{3}\left(\frac{I_p}{I_h}\right)^{\frac{3}{2}} \frac{1}{E_0 f(t)}\right], \tag{7}$$

where $I_p$ and $I_h$ represent the ionization potentials of water $1b_1$ and hydrogen $\mathcal{E}_0$ is the amplitude of the electric field while *f(t)* accounts for its time dependence. Spatially, the electrons are assumed to be emitted in a cylinder with its diameter corresponding to the laser beam's estimated waist diameter of about 50 μm and the axial dimension set to 11 cm well above the Rayleigh length of the laser beam. According to classical transport theory, the electron trajectory in an electric field can be described by a differential equation (in Hartree atomic units) of the form:

$$\ddot{x} = \dot{v} = \frac{d^2 x}{dt^2} = a = -\mathrm{E}, \tag{8}$$

with the position vector *x*, the velocity vector *v*, the acceleration vector *a*, the electric field vector $\varepsilon$ and the time *t*. The equation of motion can be solved numerically by applying a so-called Leapfrog algorithm, where the calculation of the position and the velocity happens at interleaved time points which have a distance of half the time step unit *dt*. For integer time steps *dt* with index *i*, it can be expressed as:

$$x_{i+1} = x_i + v_i dt + \frac{1}{2} a_i dt^2 \tag{9}$$

and

$$v_{i+1} = v_i + \frac{1}{2}(a_i + a_{i+1}) dt. \tag{10}$$

At each time step there is a certain possibility for a scattering event to take place, which is derived from elastic and inelastic scattering cross section. For an assumed ambient water pressure of 1 mbar however, scattering events of electrons with surrounding water molecules during the excursion in the laser electric field can be neglected.

**Electron acceleration due to space charge**

After a time of about 11 fs, the influence of the amplitude of the laser electric field can be neglected and the space charge effect becomes the dominant acceleration mechanism. Due to their large density, the free electrons strongly repel each other and are therefore accelerated. However, this acceleration is partially suppressed by the positively charged water ions originating from the ionization process. To estimate the electric field generated by the positively charged ions, a mean field approach has been chosen. Since the positively charged ions are much heavier than the electrons, they can be assumed to remain stuck at their initial position throughout the whole acceleration process. Therefore, a homogeneously charged cylinder of water ions is assumed. In order to reduce computational effort the space within and around this cylinder is divided into segments. The electrons from the first simulation are assigned to these segments and their kinematic features mass, charge, position and speed are averaged. The introduction of these "pseudo-electrons" drastically reduces the number of simulated particles and therefore the spent calculation time.

**Funding:** This research was supported by the Deutsche Forschungsgemeinschaft (DFG) via the Cluster of Excellence "e-conversion" EXC 2089/1-390776260.

**Data and materials availability:** All data needed to evaluate the conclusions in the paper are present in the paper and/or the Supplementary Materials. Additional data related to this paper may be requested from the authors.

**Figures and Tables**

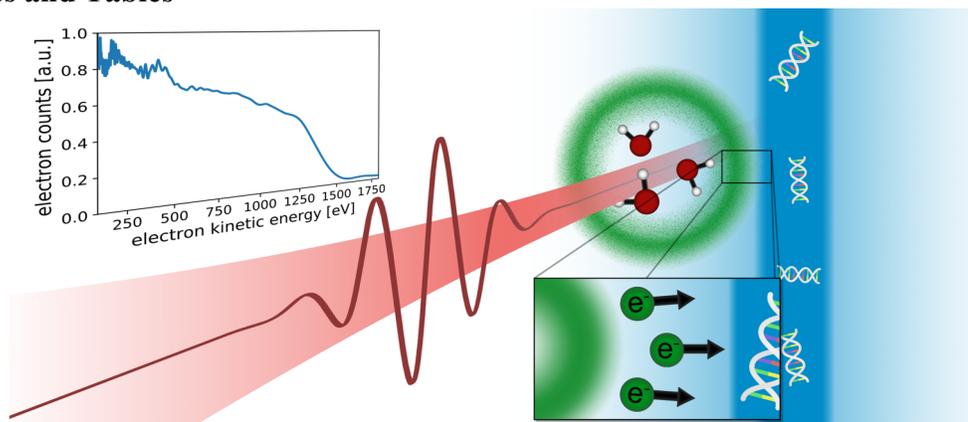

**Fig. 1. Novel scheme for ultrafast studies of electron impact in liquid water.** An ultrashort laser pulse (red) generates a cloud of free electrons in the surrounding gas phase of a liquid microjet (blue). The charged cloud (green) expands due to the repelling Coulomb force. Thereby, the outermost electrons experience the highest acceleration creating a shell of high-energy electrons followed by a trail of slower ones. Discriminating between fast and slow electrons allows for exploitation of this ultrashort bunch of fast electrons. These electrons will eventually hit the liquid microjet and can be utilized to examine dissolved organic molecules, e.g., DNA, in their natural environment.

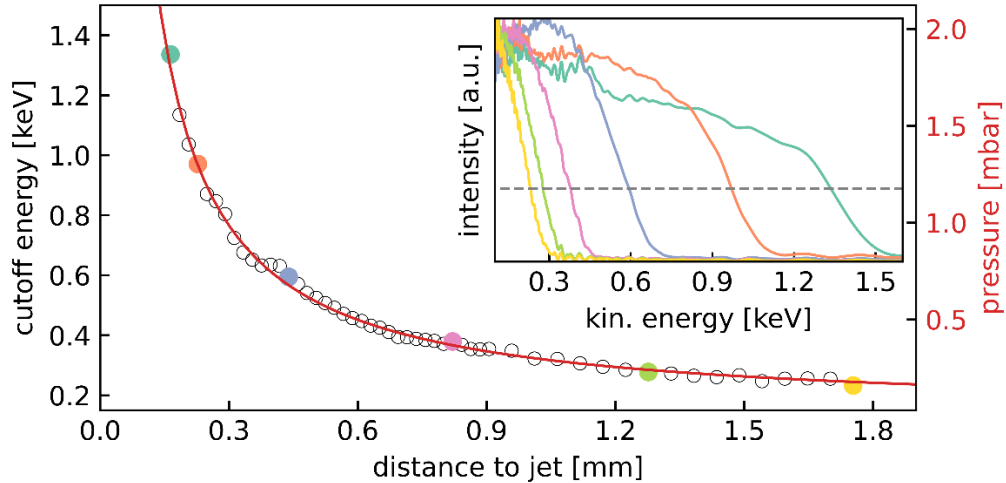

**Fig. 2. Cutoff kinetic energy versus distance between laser focus and liquid microjet.** Circles show experimental data for the cutoff energy (left scale) and determined gas pressure (right scale). The inlay shows the corresponding experimental photoelectron spectra. The grey dashed line in the inlay marks the cutoff threshold. The red line shows the radial pressure dependence as given by equation (2).

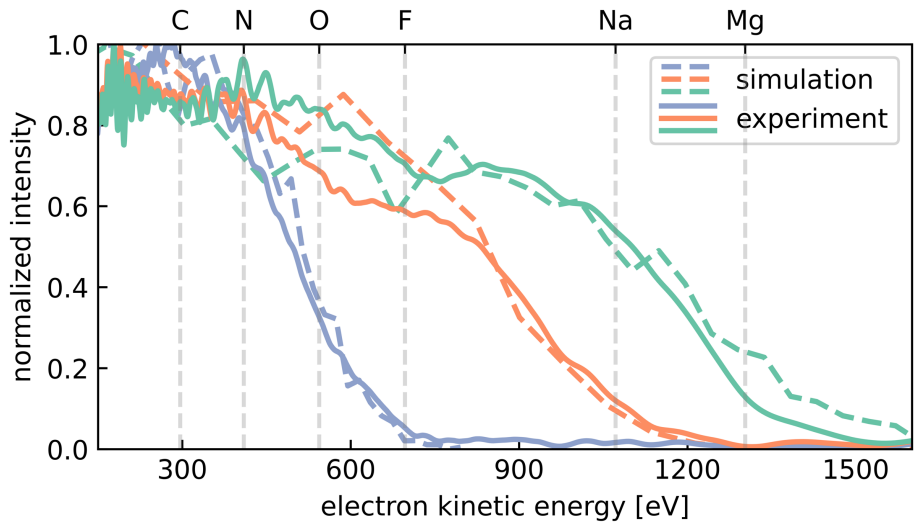

**Fig. 3. Comparison of the experimental electron spectra and simulated results.** The solid lines represent the experimental data for laser intensities of approximately $2.6\times10^{16}$ W/cm$^2$ (blue), $4.5\times10^{16}$ W/cm$^2$ (orange), and $5.3\times10^{16}$ W/cm$^2$ (green). The simulated spectra are represented as dashed lines in corresponding colors. In the simulations electron clouds of 300 (blue), 500 (orange) and 700 (green) million electrons have been assumed. To classify the cutoff energies achieved, the K-shell binding energies of selected elements are plotted on the upper abscissa.

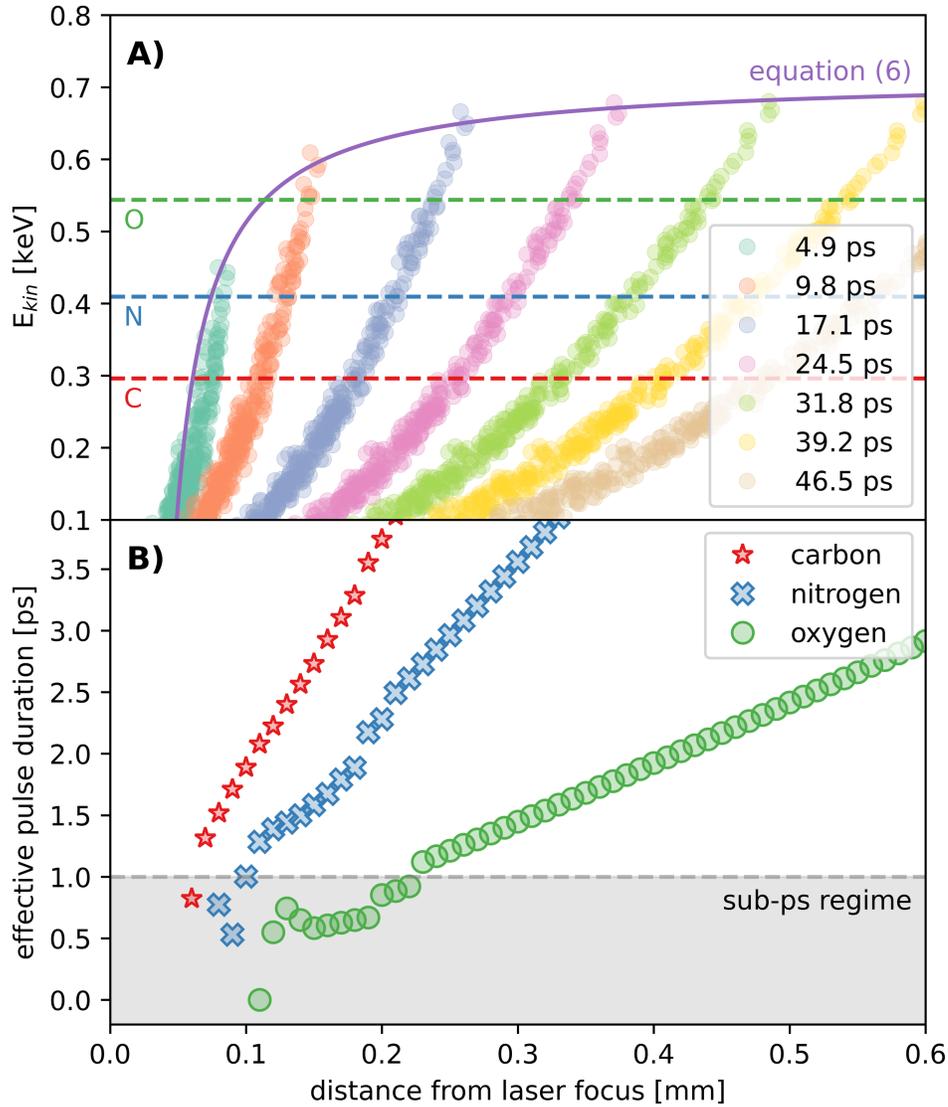

**Fig. 4. Temporal and spatial evolution of the simulated electron cloud.** (**A**) Electron kinetic energy distributions into the direction of the detector for different instants in time from 5 ps to 50 ps. The solid purple line represents the highest kinetic energy over time and arises from fitting equation (6). The dashed lines mark the energy thresholds for K-shell ionization in carbon (red), nitrogen (blue), and oxygen (green). (**B**) The full width at half maximum of the effective electron pulse duration versus its distance to the origin of the electron cloud for carbon (red stars), nitrogen (blue crosses), and oxygen (green circles). For calculation of the effective pulse duration only electrons with a kinetic energy above the threshold for the associated K-shell ionization are taken into account.

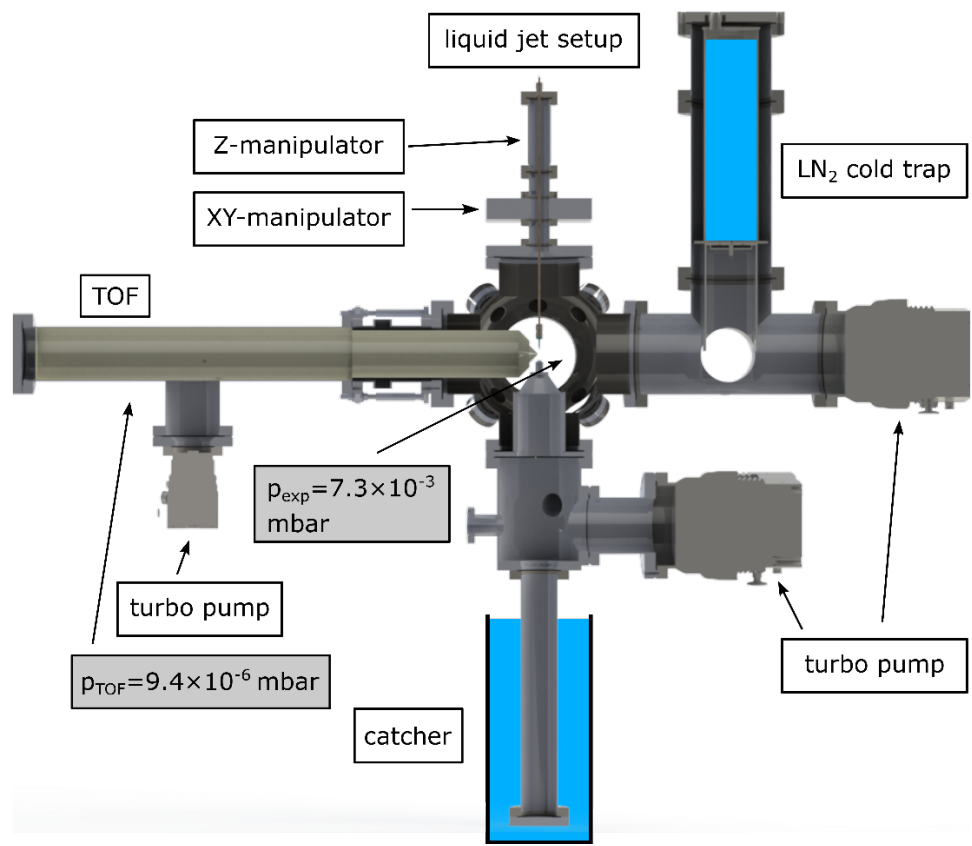

**Fig. S1. Sectional view of the vacuum system.** Two turbo pumps as well as two liquid nitrogen cooling traps are used to maintain a backing pressure of $7.3\times10^{-3}$ mbar in the experimental chamber. The small aperture of the time-of-flight spectrometer allowed another turbo pump to maintain a pressure of $9.4\times10^{-6}$ mbar inside the electron drift tube.

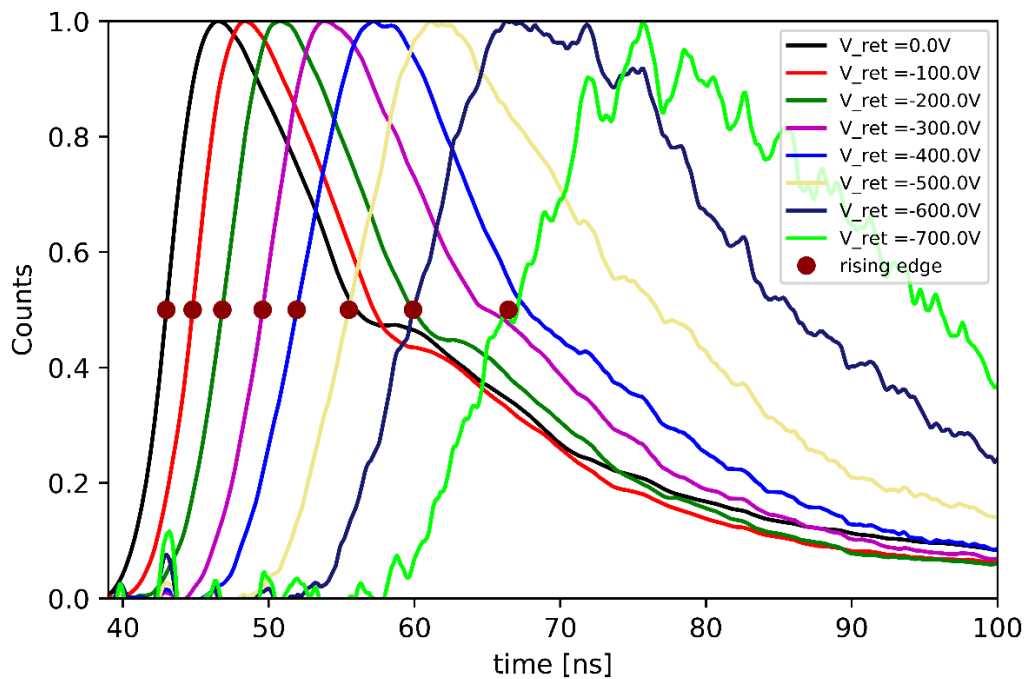

**Fig. S2. Retardation series for calibration of the time-of-flight spectrometer.** For a defined alignment the time-of-flight spectrometer drift tube has been put to a series of potentials and the arrival time of the rising edge of the electron spectrum relative to the photon prompt has been evaluated. From the link between the known applied voltage and measured arrival time the time-to-energy relation can be calculated.